\documentclass[twocolumn,twoside,prd,floatfix,letterpaper]{revtex4}

\usepackage{graphicx}
\usepackage{amsmath}
\usepackage{amssymb}
\usepackage{fancyhdr}
\usepackage{journals}
\usepackage{color}
\usepackage[colorlinks,hyperindex]{hyperref}

\def \pmiss {{\,/\!\!\!p}}

\def \figuresize {3.3in}

\newcommand {\isClickable}[1]{{}}

\def \Bard {{\sc Bard}}
\def \Quaero {{\sc Quaero}}
\def \pythia {{\sc Pythia}}
\def \Sleuth {{\sc Sleuth}}
\def \TurboSim {{\sc TurboSim}}

\def \Experiment {{\sc H1}}

\def \MadEvent {{\sc MadEvent}}
\def \MadGraph {{\sc MadGraph}}
\def \DZero {{D\O}}
\def \SM {{\ensuremath{\text{SM}}}}
\def \D {{\ensuremath{\cal D}}}
\def \H {{\ensuremath{\cal H}}}
\def \L {{\ensuremath{\cal L}}}
\def \SumPt {{\ensuremath{\sum{p_T}}}}
\def \mAll {{\ensuremath{m_\text{all}}}}

\begin{document}


\title{\Quaero@\Experiment: An Interface to High-$p_T$ HERA Event Data}
\author{Sascha Caron}
\homepage{http://hep.uni-freiburg.de/~scaron/}
\email{scaron@physik.uni-freiburg.de}
\affiliation{Physikalisches Institut, Universit\"at Freiburg, Germany}
\author{Bruce Knuteson}
\homepage{http://mit.fnal.gov/~knuteson/}
\email{knuteson@mit.edu}
\affiliation{Massachusetts Institute of Technology}

\date{\today}

\begin{abstract}
Distributions from high-$p_T$ HERA event data analyzed in a general search for new physics at \Experiment\ have been incorporated into \Quaero, an algorithm designed to automate tests of specific hypotheses with high energy collider data.  The use of \Quaero@\Experiment\ to search for leptoquarks, R-parity violating supersymmetry, and excited quarks provides examples to develop intuition for the algorithm's performance.  

\end{abstract}

\maketitle
\tableofcontents


\section{Introduction}

The publication of most searches for new physics takes the form of an exclusion contour in a two-dimensional model parameter space, with alternative interpretations typically requiring a re-analysis of the data.  Publishing the data themselves (either in raw form or as 4-vectors) is unsatisfactory, because substantial expert knowledge is required for an accurate analysis.  This article continues a recently introduced paradigm of publishing frontier energy collider data via an interface that can be used to quickly test any specific hypothesis against collider event data, with the analysis performed by an algorithm that encapsulates expert knowledge of the experiment.  In this way exclusion contours (or discovery regions) can be produced on demand.  This type of interface can be used with well understood data to test models motivated either by new theoretical insights or from the consideration of subsequently collected data.  Such an interface can also be used with freshly collected data in the exploratory process of fitting together an underlying physical interpretation for observed discrepancies while exploring in parallel more mundane experimental explanations.

The \Experiment\ detector has recorded collisions of electrons and protons at center of mass energies of 301 and 319~GeV in runs of the HERA collider between 1992 and 2000.  These data have been used to understand a wide spectrum of physics, from the internal structure of the proton to tests of exotic physics at the electroweak scale.  The analysis of HERA-I data has resulted in over one hundred publications in internationally recognized journals, and a general analysis of all high $p_T$ final states has recently been performed in the \Experiment\ General Search~\cite{H1GeneralSearch:Aktas:2004pz}.  

This article describes the interpretation of the distributions published by the H1 Collaboration in a general search for new physics using the \Quaero\ framework.
\Quaero\ was used previously by the \DZero\ collaboration to automate the optimization of searches for new physics in the Tevatron Run I data~\cite{QuaeroPRL:Abazov:2001ny}.
The H1 General Search is reviewed briefly in Sec.~\ref{sec:GeneralSearch}, with Sec.~\ref{sec:TurboSimH1} covering \Quaero's knowledge of the \Experiment\ detector response.  Section~\ref{sec:Quaero} briefly reviews the \Quaero\ algorithm.  Section~\ref{sec:Examples} contains the results of several analyses that have been performed using \Quaero@\Experiment, allowing a comparison to previous results.  A summary is given in Sec.~\ref{sec:Summary}.



\section{General Search}
\label{sec:GeneralSearch}

The \Experiment\ General Search has been published in Ref.~\cite{H1GeneralSearch:Aktas:2004pz}.  
This search, briefly described below, investigates events with high-$p_T$ objects (electrons, muons, jets, photons, and the presence of missing transverse energy) produced in $ep$ collisions at HERA. The histograms published by H1 (the invariant masses and the sums of the transverse momenta for high-$p_T$ events) are used as input to the \Quaero\ algorithm in the studies described in this paper.
This section briefly reviews the elements of this general search that have been incorporated into \Quaero@\Experiment.

A detailed description of the \Experiment\ detector can be found in Refs.~\cite{Abt:h1,Abt:1996xv}.  
The main trigger for events with high transverse momentum is provided by the liquid argon calorimeter. The trigger efficiency is close to $100\%$ for events containing an electron or photon with transverse momentum greater than $20$~GeV, $90\%$ for events containing one or more jets with $p_T>20$~GeV or with missing transverse momentum ($\pmiss_T$) greater than 20 GeV, and about $70\%$ for di-muon events~\cite{H1GeneralSearch:Aktas:2004pz}.


The \Experiment\ data available within \Quaero\ correspond to
\begin{itemize}
\item 36.4~pb$^{-1}$ of 27.5~GeV positrons on 820~GeV protons, at a center of mass energy of 301~GeV;
\item 13.8~pb$^{-1}$ of 27.5~GeV electrons on 920~GeV protons, at a center of mass energy of 319~GeV; and
\item 66.4~pb$^{-1}$ of 27.5~GeV positrons on 920~GeV protons, at a center of mass energy of 319~GeV.
\end{itemize}

Standard object identification criteria are used to define electrons ($e$), muons ($\mu$), photons ($\gamma$), and jets ($j$) ~\cite{H1GeneralSearch:Aktas:2004pz}.  All identified objects are required to have $p_T>20$~GeV and $10^\circ < \theta < 140^\circ$.  The charge of leptons is not distinguished; all leptons are taken to have positive charge.  No attempt is made to identify jets containing heavy flavor.  Events containing fewer than two of these objects are discarded.  All objects are required to be isolated by a minimum distance of one unit in the $\eta$--$\phi$ plane.  A neutrino object is defined for missing transverse momentum above $20$~GeV.  

The experimental sources of systematic error affecting the modeling of these data are identical to those considered in Ref.~\cite{H1GeneralSearch:Aktas:2004pz}.  Although \Quaero\ allows a full specification of correlated uncertainties, all \Experiment\ sources of systematic error are treated as uncorrelated.



Several Monte Carlo event generators are combined to estimate dominant Standard Model processes~\cite{H1GeneralSearch:Aktas:2004pz}.  These generated events serve as the reference model to which hypotheses presented to \Quaero\ are compared.  Here and below ``Standard Model,'' ``background,'' and ``reference model'' are used interchangeably.

\section{\TurboSim@\Experiment}
\label{sec:TurboSimH1}
\begin{figure*}
\begin{verbatim}
 1   ep->eMJX   240390.1795517551   2.98   e+   67.111   0.536   61.2   ;  ->  e+   72.969  0.531  61.37  ; 
 2   ep->eMJX   268953.1873600019  18.04   e+   16.555  -0.038   47.21  ;  ->  j    18.955  0.003  46.05  ; 
 3   ep->vMJX   271676.1791070467   5.46   j    86.379   1.689   50.96  ;  ->  j    81.975  1.686  50.95  ; 
 4   ep->eMJX   272317.1802026211  29.48   j    46.449   1.464   -7.56  ;  ->  j    27.835  1.896 -21.23  
                                                                               j    23.815  1.008  10.42  ; 
 5   ep->eMJX   278301.812519633   -5.72   j    15.606   0.225 -153.94  ;  ->  ; 
 6   ep->WX     257637.101591     -10.73   j    38.434   0.194   17.88  ;  ->  mu+  36.603  0.193  17.66  ; 
 7   ep->llX    260716.176783      -3.16   e+   11.556  -0.72   -64.11   
                                           mu+   9.506  -0.777  -59.12  ;  ->  j    22.364 -0.81  -60.98  ; 
 8   ep->eMJX   278996.533799894   -9.13   e+   30.031   0.004   12.62   
                                           ph   16.402  -0.055   10.59  ;  ->  e+   46.904 -0.028  10.96  ; 
 9   ep->llX    253700.100976       3.13   mu+  24.717   1.161  -27.75  ;  ->  ; 
10   ep->llX    256339.102973      -5.47   mu+  47.093   2.045   54.12  ;  ->  mu+  54.087  2.041  54.09  ; 
\end{verbatim}
\caption{Ten sample lines in the \TurboSim@\Experiment\ lookup table, chosen to illustrate \TurboSim's handling of interesting cases.  Each line begins with the event's type, run and event number, and vertex position.  To the left of the arrow (``{\tt ->}'') is a list of nearby parton-level objects; to the right of the arrow is a list of corresponding reconstructed-level objects.  The first line shows a positron correctly identified as a positron, while the second line shows a positron that has not been correctly identified.  The third line shows a nicely reconstructed jet; in the fourth line the jet has been split into two; in the fifth line the jet is either not reconstructed or reconstructed with $p_T<20$~GeV.  The sixth line shows a jet identified as a muon.  The seventh line shows a nearby positron and muon that have been reconstructed as one jet; the eighth line shows a positron with a photon radiated sufficiently nearby that the two are agglomerated into a single reconstructed-level positron.  The ninth and tenth lines show muons that are missed and reconstructed, respectively.}
\label{fig:TurboSimH1Table}
\end{figure*}

To keep \Quaero\ fast and standalone, a fast detector simulation algorithm (\TurboSim@H1) is built in accordance with the H1 detector simulation.  It is based on a large lookup table of one half million lines mapping particle-level objects to objects reconstructed in the detector. 
Sample lines in this table are shown in Fig.~\ref{fig:TurboSimH1Table}.  The total table is roughly 100~MB, and as such can be read into memory and searched as a multivariate binary tree.  The resulting simulation runs at roughly 10~ms per event.  


Particle efficiencies are handled through lines in the \TurboSim@\Experiment\ table that map a parton level object to no reconstructed level object.  Misidentification probabilities are handled through lines that map a particle-level object to a reconstructed-level object of a different type.  The merging and overlap of particles is handled by configurations in the table that map two or three particle-level objects to zero or more reconstructed-level objects.  

Validation of \TurboSim@H1 has been performed by running an independent sample of one million events through both the H1 full simulation and \TurboSim@H1. The event classification and the kinematic distributions of the events from the two simulation chains are compared using a Kolmogorov-Smirnov (KS) test.  \TurboSim@H1 has been found in agreement with the full simulation of H1.

\section{\Quaero}
\label{sec:Quaero}

\begin{figure}
\href{http://mit.fnal.gov/Quaero}{\includegraphics[width=\figuresize]{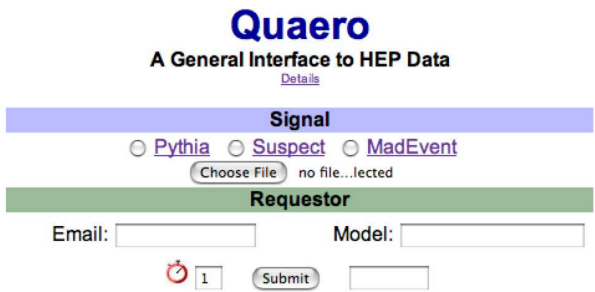}}
\caption{The \Quaero\ interface, designed for HERA-I, LEP\,2, Tevatron II, and the future LHC.  A new hypothesis can be provided as commands to one of several event generators.  The result returned quantifies the extent to which the data (dis)favors the new hypothesis relative to the Standard Model.}
\label{fig:QuaeroWebPage}
\end{figure}

\Quaero\ provides a convenient interface to the understanding represented by high energy collider data, backgrounds, and detector response.  This interface is designed to facilitate the test of any specific hypothesis against such data.  The \Quaero\ web page, shown in Fig.~\ref{fig:QuaeroWebPage}, is available online at Ref.~\cite{QuaeroURL}.
  
A physicist wishing to test her hypothesis against \Experiment\ data will provide her hypothesis in the form of commands to one of the built-in event generators~\cite{Pythia:Sjostrand:2000wi,Suspect:Djouadi:2002ze,MadEvent:Maltoni:2002qb2}.  \Quaero\ uses the specified event generator to generate signal events corresponding to $e^+p$ collisions at 301~GeV, $e^-p$ collisions at 319~GeV, and $e^+p$ collisions at 319~GeV.  The response of the \Experiment\ detector to these events is simulated using \TurboSim@\Experiment.  The output of \TurboSim@\Experiment\ is a text file containing simulated events, normalized such that the total number of expected signal events produced at \Experiment\ is equal to the sum of all event weights in the file.  

Three distinct samples of events exist at this point: the data \D; the Standard Model prediction \SM; and the hypothesis \H, which is the sum of included Standard Model processes and the physicist's signal.  Each sample of events is partitioned into exclusive final states, categorized by reconstructed objects with $p_T>20$~GeV.  In each exclusive final state, a pre-defined list of 
two variables --- 
the summed scalar transverse momentum (\SumPt) and the invariant mass of all objects (\mAll)
--- are ranked according to the difference between the Standard Model prediction and the physicist's hypothesis \H.  
The variable showing the most difference is used~\footnote{Although the \Quaero\ algorithm is able to choose among many relevant kinematic variables, limitation to the information content of Ref.~\cite{H1GeneralSearch:Aktas:2004pz} requires restriction to either the single variable \SumPt\ or \mAll.}.
 
In this variable space, densities are estimated from the Monte Carlo events predicted by \SM\ and \H.  These densities are used to define a discriminant, which is binned to distinguish \SM\ from \H.  The likelihood ratio $\L = p(\D|\H)/p(\D|\SM)$ is determined using this binning, and systematic errors are integrated numerically.  

The result returned by \Quaero\ is the decimal logarithm of this likelihood ratio.  The measurement of model parameters using \Quaero\ is easily accomplished by graphing $\log_{10}{\L}$ as a function of varied parameter values, with multiple \Quaero\ submissions.  Distributions of the Standard Model prediction \SM, the prediction of the physicist's hypothesis \H, and the \Experiment\ data \D\ in the most relevant variable in each of the most relevant final states are returned to the querying physicist in an email along with her result.  

Further details of the \Quaero\ algorithm are provided in Ref.~\cite{QuaeroAlgorithm}.

\section{Examples}
\label{sec:Examples}

\Quaero\ has been used to test models that have previously been considered at \Experiment, in order to benchmark \Quaero's sensitivity, and to test models that have not yet been considered at \Experiment, in order to see how \Quaero\ performs on novel searches for new physics.  These examples provide intuition for the \Quaero\ algorithm:  its strengths, and its limitations.  

A rough, non-rigorous, but nonetheless useful comparison of the sensitivity of \Quaero's results (which take the form of the decimal logarithm of a likelihood ratio) with previous analyses (which typically take the form of 95\% confidence level exclusion limits) can be made by comparing $\log_{10}{\L}=-1$ with the 95\% confidence level exclusion limit~\footnote{Consider a model with an overall cross section $\sigma$ as its only free parameter, and a flat Bayesian prior on $\sigma$.  After performing an analysis, suppose the posterior cross section distribution $p(\sigma)$ is a gaussian centered at zero, with the restriction $\sigma>0$.  Then models with $\sigma$ greater than two standard deviations away from zero are excluded at a confidence level of 95\%.  These correspond to models with $\log_{10}{\L}<\log_{10}{\exp\left(-\frac{1}{2}2^2\right)}=-0.87$.  Suppose instead the posterior cross section distribution $p(\sigma)$ is a decreasing exponential, with the restriction $\sigma>0$.  Then models with $\sigma$ greater than three times the decay length of the exponential are excluded at a confidence level of 95\%.  These correspond to models with $\log_{10}{\L}<\log_{10}{\exp(-3)}=-1.3$.  In this article $\log_{10}{\L}=-1$ is highlighted as a rough and convenient choice for the purpose of building intution when comparing with previous results.}.  The decimal logarithm of the likelihood returned by \Quaero\ can be converted into exclusion limits, measurements with errors, or (potentially) statements of discovery, perhaps with multiple \Quaero\ submissions.

The examples considered in this section include a search for scalar leptoquarks, 
R-parity violating supersymmetry, and an excited quark.

\subsection{Leptoquarks}
\label{sec:Leptoquark}

\begin{figure}
\begin{verbatim}
###############################################
           Name  Anti  Spin  Mass  Width  Color 
           xxxx  xxxx  sfv   GeV    GeV    sto  

PARTICLE    lq    lq~   s    301    0.54    t
###############################################
           Interacting particles    Coupling
INTERACTION   u       e-      lq      0.3
INTERACTION   e-      u       lq~     0.3
INTERACTION   g       lq~     lq      1.22
###############################################
\end{verbatim}
\caption{\MadEvent\ input given to \Quaero\ to specify the Lagrangian terms in Eq.~\ref{eqn:Lagrangian_lq}.  The line beginning with {\tt PARTICLE} introduces into the theory a color triplet scalar leptoquark with mass 301~GeV.  The three lines beginning with {\tt INTERACTION} introduce new vertices and specify the coupling strength at each vertex.}
\label{fig:MadEventInput_lq}
\end{figure}

\Quaero@\Experiment\ is first used to search for leptoquarks, particles possessing both lepton and baryon quantum numbers that arise naturally in Grand Unified Theories.  Attention is restricted to a scalar leptoquark coupling to a positron and an up quark.  The coupling $\lambda$ of the LQ-$e$-$u$ vertex and the leptoquark mass $m_{\text{LQ}}$ are allowed to vary.  The interaction Lagrangian is assumed to be of the form

\begin{eqnarray}
{\cal L} & = & \lambda \, \text{LQ} \, \bar{u}_R \, e_L + {\text{h.c.}} \nonumber \\
         &   & + i g_s G_\mu^* ( \text{LQ}^* \overleftrightarrow{\partial}^\mu \text{LQ} ),
\label{eqn:Lagrangian_lq}
\end{eqnarray}
where \text{LQ} is a scalar leptoquark field; $\bar{u}_R$ and $e_L$ represent a right-handed anti-up quark and left-handed electron; $G_\mu$ is the gluon field; and $g_s=\sqrt{4\pi\alpha_s} \approx 1.2$.
The \MadEvent~\cite{MadEvent:Maltoni:2002qb2} input to \Quaero\ corresponding to these Lagrangian terms is shown in Fig.~\ref{fig:MadEventInput_lq}.

The likelihood ratio \Quaero\ returns should be interpreted as an update of betting odds for or against the provided hypothesis.  For the hypothesis of a scalar leptoquark with mass $m_{\text{LQ}}=301$~GeV, \Quaero\ finds $\log_{10}{\L}= -1.5$.  If betting odds on this hypothesis were (say) 100:1 against before looking at these data, then these data indicate those odds should be adjusted by an extra factor of $10^{-1.5} \approx 32$:1 against.  Betting odds against this hypothesis after having run this request are now 3200:1.  


\Quaero\ also returns a link to plots showing how the result has been obtained.  The cover page to these distributions contains the requester's name, email address, and the \Quaero\ request number, together with the contributions of each experiment to the final \Quaero\ result.  One plot is returned for each of the final states contributing greater than 0.1 to the decimal logarithm of the likelihood ratio.  

\begin{figure}
\href{http://mit.fnal.gov/Quaero/quaero/doc/examples/clickableExclusionPlots/h1_lq.html}{\includegraphics[width=\figuresize]{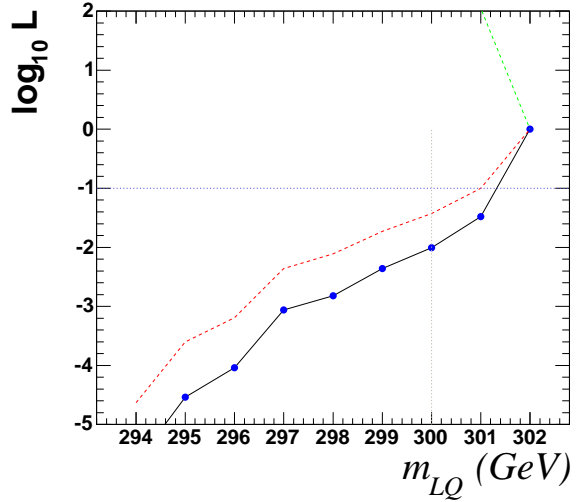}}
\caption{\Quaero's output ($\log_{10}{\L}$) for the interaction Lagrangian of Eq.~\ref{eqn:Lagrangian_lq} with $\lambda=0.3$ as a function of assumed leptoquark mass ($m_{\text{LQ}}$), shown as the solid line through filled circles.  The evidence \Quaero\ expects to provide in favor of the Standard Model \SM\ is shown as the (dark, red) dashed curve with $\log_{10}{\L}<0$; the evidence \Quaero\ expects to provide in favor of the hypothesis \H\ is shown as the (light, green) dashed curve with $\log_{10}{\L}>0$.  The result of a previous analysis excludes leptoquark masses $m_{\text{LQ}} \lesssim 300$~GeV at a confidence level of 95\%, indicated by the vertical (gray) dashed line.  
\isClickable{This exclusion plot is clickable, providing access to \Quaero's analysis of each parameter point.}
}
\label{fig:H1Lq_logL}
\end{figure}

The result of \Quaero's search for leptoquarks with interactions specified by the Lagrangian of Eq.~\ref{eqn:Lagrangian_lq} with $\lambda=0.3$ and with mass up to 302~GeV is shown in Fig.~\ref{fig:H1Lq_logL}.  In all cases \Quaero\ finds $\log_{10}{\L}\leq0$, indicating the \Experiment\ data favor the Standard Model over the provided hypotheses.  


In addition to varying the assumed leptoquark mass, the coupling strength $\lambda$,
the signal production cross section, or an overall $k$-factor can be separately specified.
An algorithm for turning this \Quaero\ request into a standard 95\% confidence level exclusion limit involves submitting several requests assuming a range of leptoquark masses and a range of different cross sections at each mass, and computing the 95\% confidence level cross section limit at each mass assuming a prior distribution for that cross section.  Masses for which the 95\% confidence level cross section limit is smaller than the theoretical cross section are said to be excluded at 95\% confidence.

Intuitively, the 95\% confidence exclusion region corresponds to a region where the signal hypothesis is moderately disfavored by the data.  If the threshold is chosen to be 10:1 against ($\log_{10}{\L}=-1$), then leptoquarks of mass less than 301~GeV are excluded.  This result is in accord with the result of a previous analysis of this signal, described in Ref.~\cite{previousH1LeptoquarkAnalysis:Aktas:2005pr}, which derives a comparable limit for leptoquarks of this type.  

\begin{figure}
\begin{tabular}{cc}
\href{http://mit.fnal.gov/Quaero/quaero/doc/examples/clickableExclusionPlots/d0_lq_sue+.html}
{\includegraphics[width=1.75in]{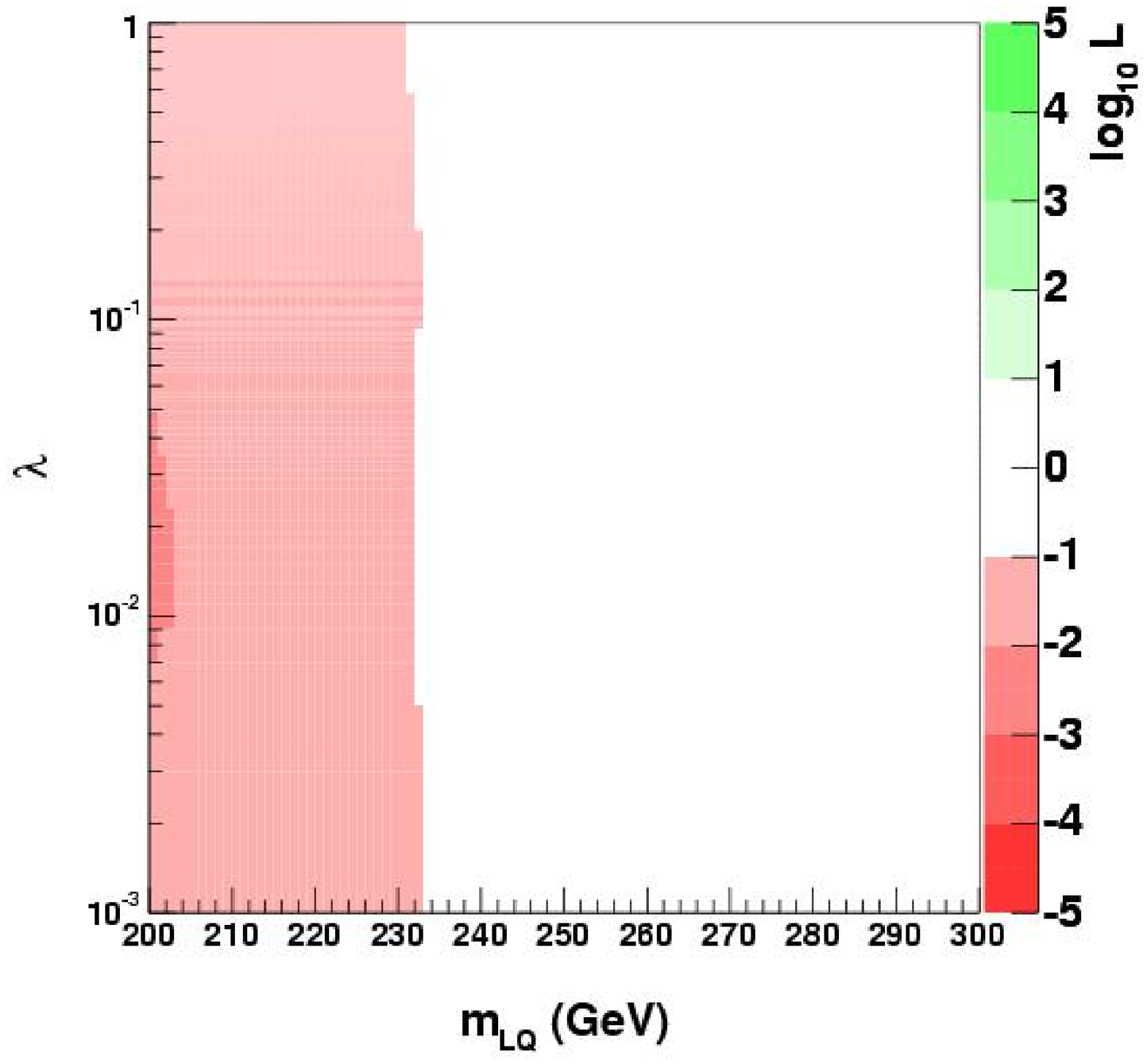}} &
\href{http://mit.fnal.gov/Quaero/quaero/doc/examples/clickableExclusionPlots/h1_lq_sue+.html}
{\includegraphics[width=1.75in]{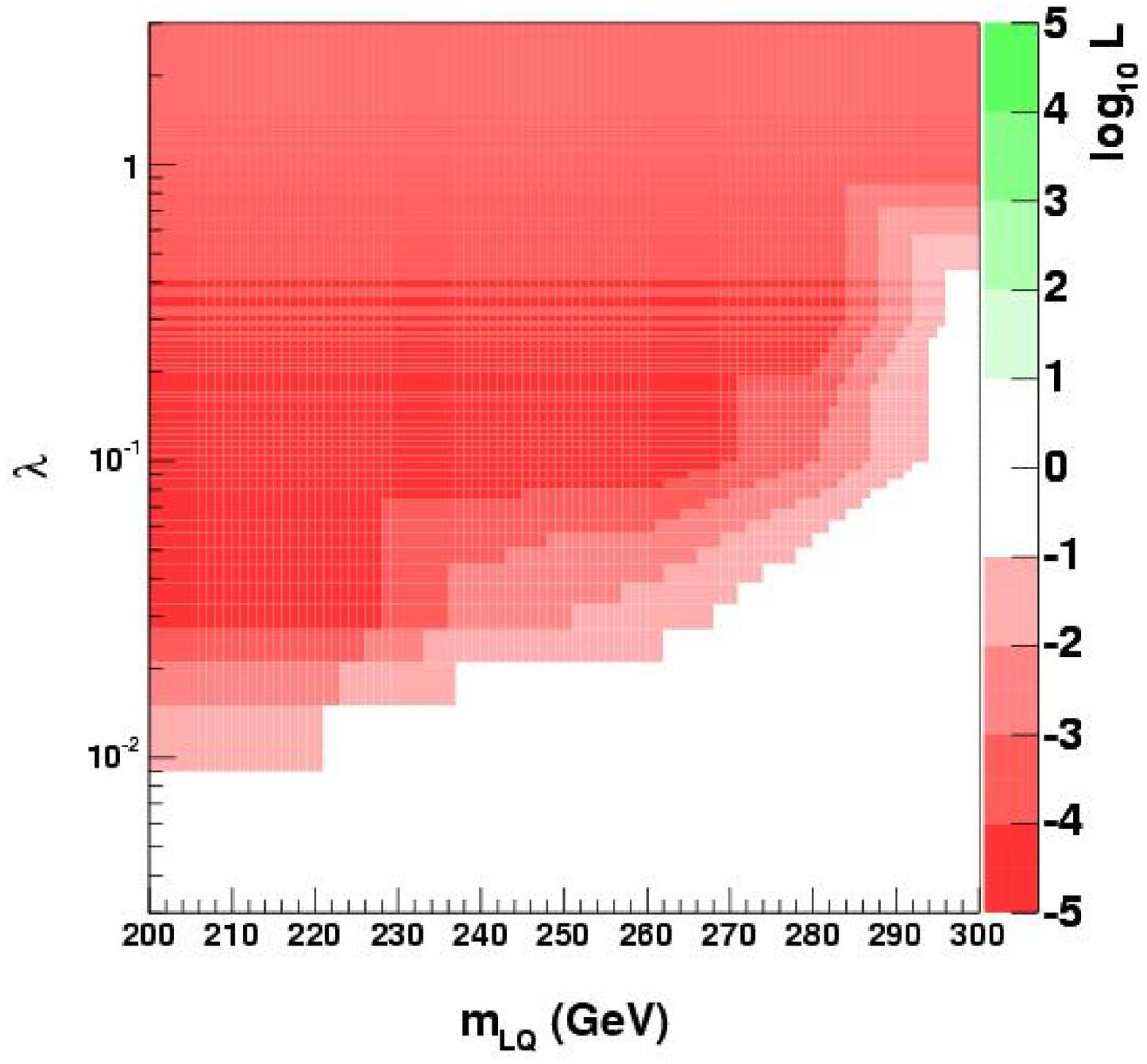}} \\
\end{tabular}
\caption{\Quaero's log likelihood ratio as a function of the coupling $\lambda$ and leptoquark mass $m_{\text{LQ}}$, in the scenario defined by the additional interaction terms of Eq.~\ref{eqn:Lagrangian_lq}.  Shown separately are results from \Quaero\ using data from D\O\ Tevatron Run I (left) and using data from H1 HERA Run I (right).  All shaded area corresponds to $\log_{10}{L}<0$.  
\isClickable{These exclusion plots are clickable, providing access to \Quaero's analysis of each parameter point.}
}
\label{fig:separate_lq_logL_2d}
\end{figure}

\begin{figure}
\href{http://mit.fnal.gov/Quaero/quaero/doc/examples/clickableExclusionPlots/all_lq_sue+.html}
{\includegraphics[width=\figuresize]{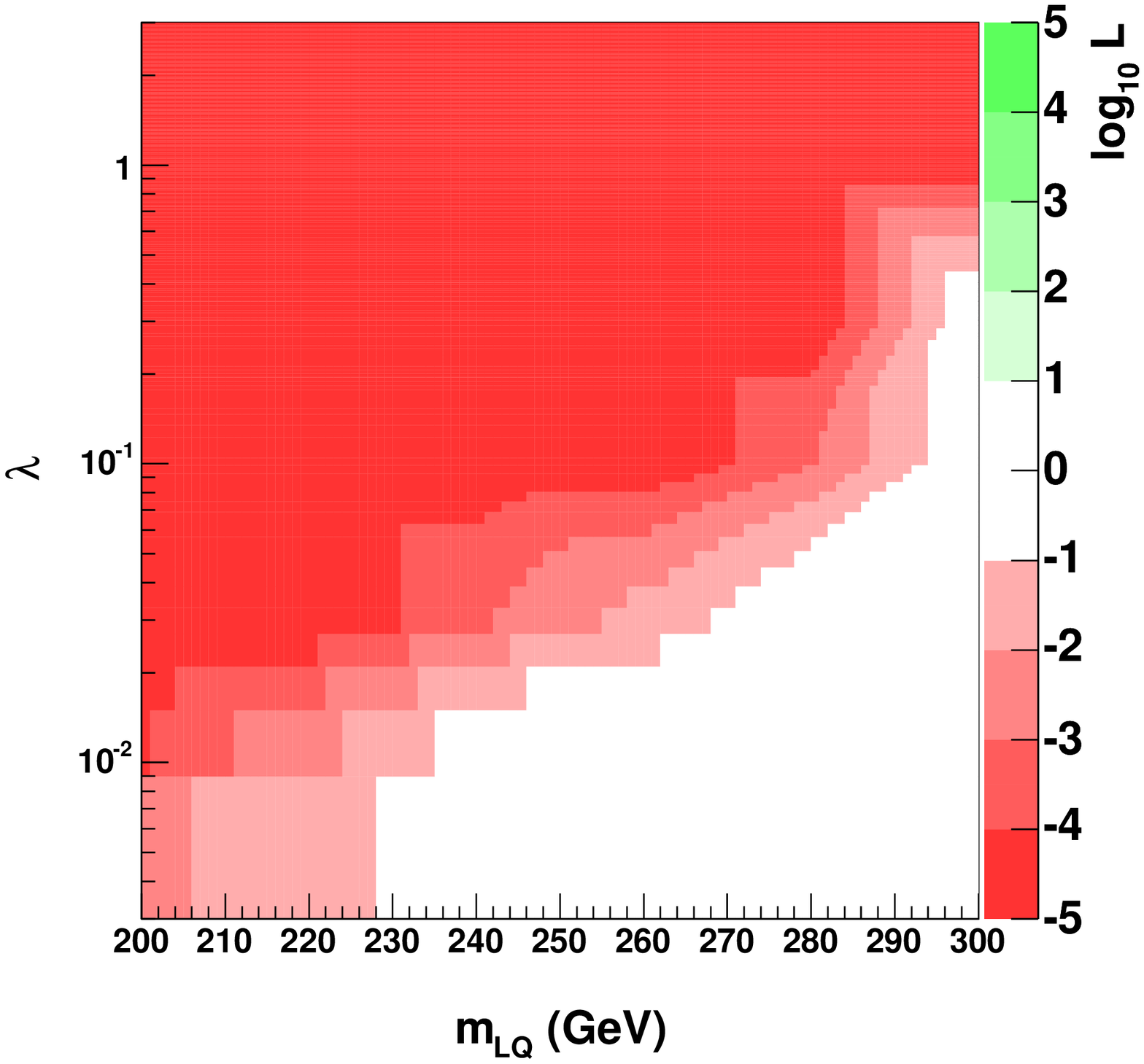}}
\caption{\Quaero's log likelihood ratio as a function of the coupling $\lambda$ and leptoquark mass $m_{\text{LQ}}$, in the scenario defined by the additional interaction terms of Eq.~\ref{eqn:Lagrangian_lq}, combining data from D\O\ Tevatron Run I and H1 HERA Run I.  All shaded area corresponds to $\log_{10}{L}<0$.  
\isClickable{This exclusion plot is clickable, providing access to \Quaero's analysis of each parameter point.}
}
\label{fig:all_lq_logL_2d}
\end{figure}

The subset of D\O\ Run I data made available in the first implementation of \Quaero~\cite{QuaeroPRL:Abazov:2001ny} has been incorporated into the current version of \Quaero.  Plots of \Quaero's result in the parameter plane of $\lambda$ and $m_{\text{LQ}}$ using the H1 and D\O\ data separately are shown in Fig.~\ref{fig:separate_lq_logL_2d}.  \Quaero's result using H1 and D\O\ data combined is shown in Fig.~\ref{fig:all_lq_logL_2d}.  \Quaero\ is able to make use of the Tevatron's $\lambda$-independent exclusion of leptoquarks with low mass and HERA's $\lambda$-dependent exclusion at higher masses to rule out more of the parameter space than either collider is able to on its own.  


\subsection{R-parity violating supersymmetry}
\label{sec:muonSneutrinoLsp}

\begin{figure}
\includegraphics[width=2.25in]{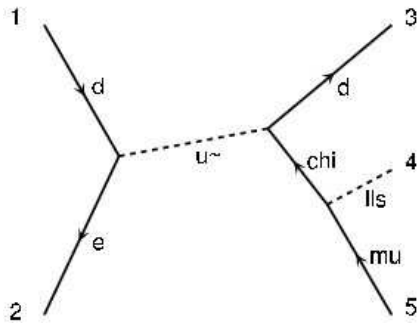}
\caption{A Feynman diagram generated by \MadGraph~\cite{MadGraph:Stelzer:1994ta} given the additional particles and interactions specified in Fig.~\ref{fig:MadEventInput_muonSneutrinoLsp}.  Here {\tt u}$\sim$ is an up squark, {\tt chi} is a chargino, and {\tt lls} is a long-lived scalar, in this case a muon sneutrino.}
\label{fig:diagram_muonSneutrinoLsp}
\end{figure}

An R-parity violating supersymmetry scenario is considered as a possible explanation for the $\mu j \nu$ events whose interestingness is quantified in Ref.~\cite{H1GeneralSearch:Aktas:2004pz}.  R-parity violating supersymmetry can lead to squark production at HERA.  If the muon sneutrino happens to be the lightest supersymmetric particle, then diagrams such as that shown in Fig.~\ref{fig:diagram_muonSneutrinoLsp} can give rise to the final state $\mu j \nu$.  This represents a scenario not yet tested by either of the HERA experiments.

\begin{figure}
\begin{verbatim}
###############################################
           Name  Anti  Spin  Mass  Width  Color 
           xxxx  xxxx  sfv   GeV    GeV    sto  

PARTICLE   usq   usq~   s    200     1      t    
PARTICLE   chi-  chi+   f    150     1      s    
###############################################
           Interacting particles    Coupling
INTERACTION   d       e-      usq     0.01
INTERACTION   chi-    d       usq~    0.05
INTERACTION   mu-     chi-    lls     0.10

INTERACTION   e-      d       usq~    0.01
INTERACTION   d       chi-    usq     0.05
INTERACTION   chi-    mu-     lls     0.10
###############################################
PARAMETER  long_lived_mass = 50
###############################################
\end{verbatim}
\caption{\MadEvent\ input given to \Quaero\ to specify a signal corresponding to the new particles and interactions that must be introduced in order for a diagram such as that shown in Fig.~\ref{fig:diagram_muonSneutrinoLsp} to occur.  The two lines beginning with {\tt PARTICLE} introduce into the theory a $u$ squark with spin 0, mass 200~GeV, width 1~GeV, and a color triplet; and a $\chi^\pm$ with spin $1/2$, mass 150~GeV, width 1~GeV, and a color singlet.  The six lines beginning with {\tt INTERACTION} introduce new vertices and specify the coupling strength at each vertex.  The particle {\tt lls} is a ``long-lived scalar'' that is implicitly added to the particle content of the theory; a long-lived fermion {\tt llf} and a long-lived vector {\tt llv} are similarly available.  The line beginning with {\tt PARAMETER} specifies the mass of the {\tt lls}.}
\label{fig:MadEventInput_muonSneutrinoLsp}
\end{figure}

\begin{figure}
\includegraphics[width=\figuresize]{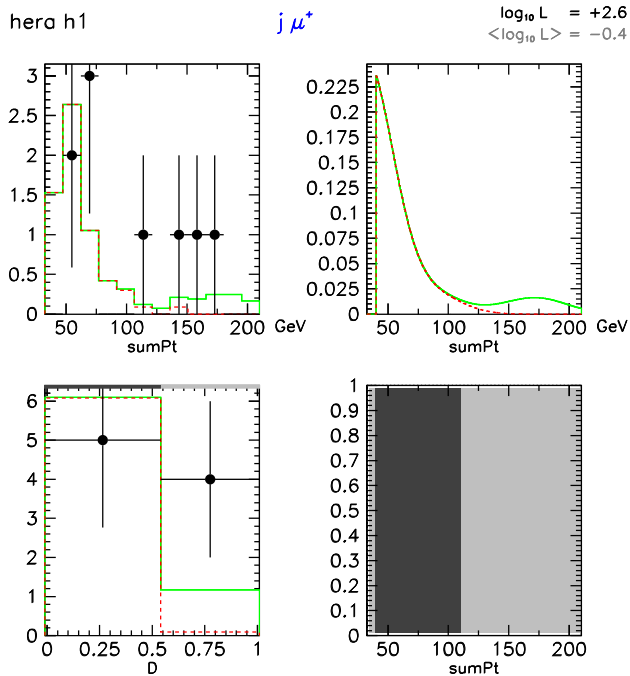}
\caption{Plots illustrating \Quaero's testing of the R-parity violating supersymmetry scenario described in the text, with the muon sneutrino playing the role of the lightest supersymmetric particle.  The contribution of the postulated signal is seen (upper left) by comparing the distributions predicted by the hypothesis \H\ (light, green) to the Standard Model \SM\ (dark, red) in the final state $j\mu^+$ in \Quaero's chosen variable {\tt sumPt} (\SumPt).  \Experiment\ data are shown as (black) filled circles.  The upper right figure shows the density estimates constructed by \Quaero\ in this variable.  Bins in the resulting discriminant are shown in the lower two figures.  Shading in the lower two plots indicates the bin $D \lesssim 0.54$ corresponds to $40 \lesssim \SumPt \lesssim 110$, and the bin $D \gtrsim 0.54$ corresponds to the signal-rich region $110 \lesssim \SumPt$.  The vertical axis in the lower right plot is without meaning, intended simply to give ``thickness'' to the shaded regions in {\tt sumPt}.  The number (black) at upper right is the log likelihood ratio $\log_{10}{\L}$ that would be found by \Quaero\ if that final state were analyzed alone, without integration over systematic errors; the number (gray) just below this is the log likelihood ratio expected if the data are drawn from the Standard Model distribution.  Systematic errors are included in the final result returned by \Quaero.
}
\label{fig:h1_muonSneutrinoLsp_expertPlots}
\end{figure}

Most new physics scenarios are specified ``top-down,'' starting with the general Lagrangian and then restricting the resulting phenomenology through a judicious choice of values of the model parameters.  In this subsection a ``bottom-up'' approach is adopted to illustrate another way in which \Quaero\ may be used to understand features of the data in terms of the underlying physical theory, in the spirit of \Bard~\cite{BardPRL:Knuteson:2006ha}.  A new physics process is guessed, together with required masses and couplings, and rigorously tested using \Quaero.  Iteration of this procedure allows a fit of parameters.

\begin{figure}
\href{http://mit.fnal.gov/Quaero/quaero/doc/examples/clickableExclusionPlots/h1_muonSneutrinoLsp.html}
{\includegraphics[width=\figuresize]{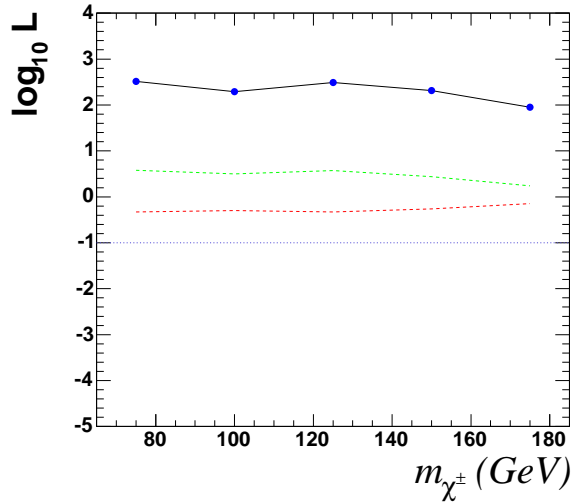}}
\caption{\Quaero's log likelihood ratio (solid curve) as a function of assumed $m_{\chi}$.  Also shown is \Quaero's expected evidence (dashed curves) as a function of this parameter, if the data are drawn from the hypothesis (light, green) or the Standard Model (dark, red).  The hypothesis provides a better fit to the data than the Standard Model for all values of $m_{\chi}$ in the range shown, corresponding to an update of betting odds of $\approx$ 300:1 in favor.  
\isClickable{This exclusion plot is clickable, providing access to \Quaero's analysis of each parameter point.}
}
\label{fig:h1_muonSneutrinoLsp_logL_1d}
\end{figure}

Ignoring the full sparticle spectrum, the new particles that must be postulated to allow the diagram shown in Fig.~\ref{fig:diagram_muonSneutrinoLsp} are a scalar color triplet ($\tilde{u}$), an electrically charged fermion color singlet ($\chi^+$), and a stable scalar without strong or electromagnetic interactions ($\tilde{\nu}_\mu$).  Three new interactions must also be postulated, corresponding to the three vertices of Fig.~\ref{fig:diagram_muonSneutrinoLsp}.

These new particles and interactions can be provided to \Quaero\ as \MadEvent\ input, shown in Fig.~\ref{fig:MadEventInput_muonSneutrinoLsp}, without specifying the complete renormalizable theory.

Taking the widths of each new particle to be small compared to experimental resolution, this scenario contains six new parameters:  three new particle masses and three new coupling strengths.   The kinematics of the $\mu j \nu$ events suggest $m_{\tilde{u}}\approx200$~GeV and $m_{\tilde{\nu}_\mu}\lesssim 50$~GeV, while bounds from LEP\,1 indicate $m_{\tilde{\nu}_\mu}>M_Z/2$, leading to the choice of $m_{\tilde{u}}=200$~GeV and $m_{\tilde{\nu}_\mu}=50$~GeV for this example.  The number of events observed by \Experiment\ in the $\mu j \nu$ final state suggest a signal production cross section of roughly $0.1$~pb.  Considering the single diagram of Fig.~\ref{fig:MadEventInput_muonSneutrinoLsp}, the three new couplings affect the phenomenology of the postulated signal only by multiplying the overall cross section by the square of their product.  The $\chi$-$d$-$\tilde{u}$ and $\mu$-$\chi$-$\tilde{\nu}_\mu$ couplings are of electroweak strength; the R-parity violating coupling $d$-$e$-$\tilde{u}$ is chosen so that the cross section of the process shown in Fig.~\ref{fig:diagram_muonSneutrinoLsp} is 0.1~pb.  The remaining parameter $m_{\chi}$ is allowed to vary between 75~GeV and 175~GeV.

Figure~\ref{fig:h1_muonSneutrinoLsp_expertPlots} shows \Quaero's analysis of this hypothesis with the choice $m_{\chi}=150$~GeV.  As expected, the final state $j\mu(\nu)$ contributes most to the final result.  \Quaero\ chooses to consider the variable \SumPt\ in this final state.  \Quaero's density estimates and choice of binning are shown in the right panes of Fig.~\ref{fig:h1_muonSneutrinoLsp_expertPlots}.  The left panes of Fig.~\ref{fig:h1_muonSneutrinoLsp_expertPlots} suggest that the data is better fit by this hypothesis than by the Standard Model alone, a conclusion \Quaero\ quantifies in its returned result of $\log_{10}{\L}=2.5$.  \footnote{\Quaero's apparent insensitivity to $m_{\chi}$ is due to restriction to the crude variables \SumPt\ and \mAll; allowed unrestricted choice of kinematic variables, \Quaero\ chooses observables that closely approximate $m_{\chi}$.  The model-independence of the variables \SumPt\ and \mAll\ are the reason these variables are used in the model-independent analysis of Ref.~\cite{H1GeneralSearch:Aktas:2004pz}, and the reason the variable \SumPt\ is used in \Sleuth~\cite{acat2003proceedings:Knuteson:2004nj,SleuthPRL:Abbott:2001ke,SleuthPRD1:Abbott:2000fb,SleuthPRD2:Abbott:2000gx}.}

\Quaero's result $\log_{10}{\L}$ as a function of assumed $m_{\chi}$ is shown in Fig.~\ref{fig:h1_muonSneutrinoLsp_logL_1d}.  Over the range shown, this hypothesis is found to fit the data better than the Standard Model alone, and the decimal logarithm of the likelihood ratio is $\log_{10}{\L} \approx 2.5$.  As mentioned previously, this likelihood ratio may be interpreted in terms of an ``update of betting odds'' for this hypothesis relative to the Standard Model.  If betting odds against this hypothesis are (say) $10^7$:1 before the \Experiment\ data are considered, these betting odds should be revised in light of these data to $\approx$ 30000:1 against.

\begin{figure}
\begin{verbatim}
###############################################
           Name  Anti  Spin  Mass  Width  Color 
           xxxx  xxxx  sfv   GeV    GeV    sto  

PARTICLE    ux    ux~   f    200    0.04    t
###############################################
       Interacting particles   Coupling
INTERACTION  u      ux     a   0.000667  dmx
INTERACTION  ux     u      a   0.000667  dmx

INTERACTION  d      ux     w-  0.00147   dmx
INTERACTION  ux     d      w+  0.00147   dmx

INTERACTION  u      ux     z   0.000825  dmx
INTERACTION  ux     u      z   0.000825  dmx

INTERACTION  ux     ux     a   0.3
INTERACTION  ux     ux     z   0.3
###############################################
\end{verbatim}
\caption{\MadEvent\ input given to \Quaero\ to specify an excited quark signal.  The symbols {\tt u} and {\tt d} denote the Standard Model up quark and down quark; {\tt ux} and {\tt dx} denote an excited up quark and an excited down quark.  The line beginning with {\tt PARTICLE} introduces into the theory an excited up quark with spin $1/2$, mass 200~GeV, width 0.04~GeV, and a color triplet.  The eight lines beginning with {\tt INTERACTION} introduce new vertices and specify the coupling strength at each vertex, with {\tt dmx} indicating a dipole moment coupling.  An excited down quark and its interactions are similarly specified.}
\label{fig:MadEventInput_excitedQuark}
\end{figure}

\begin{figure}
\includegraphics[width=\figuresize]{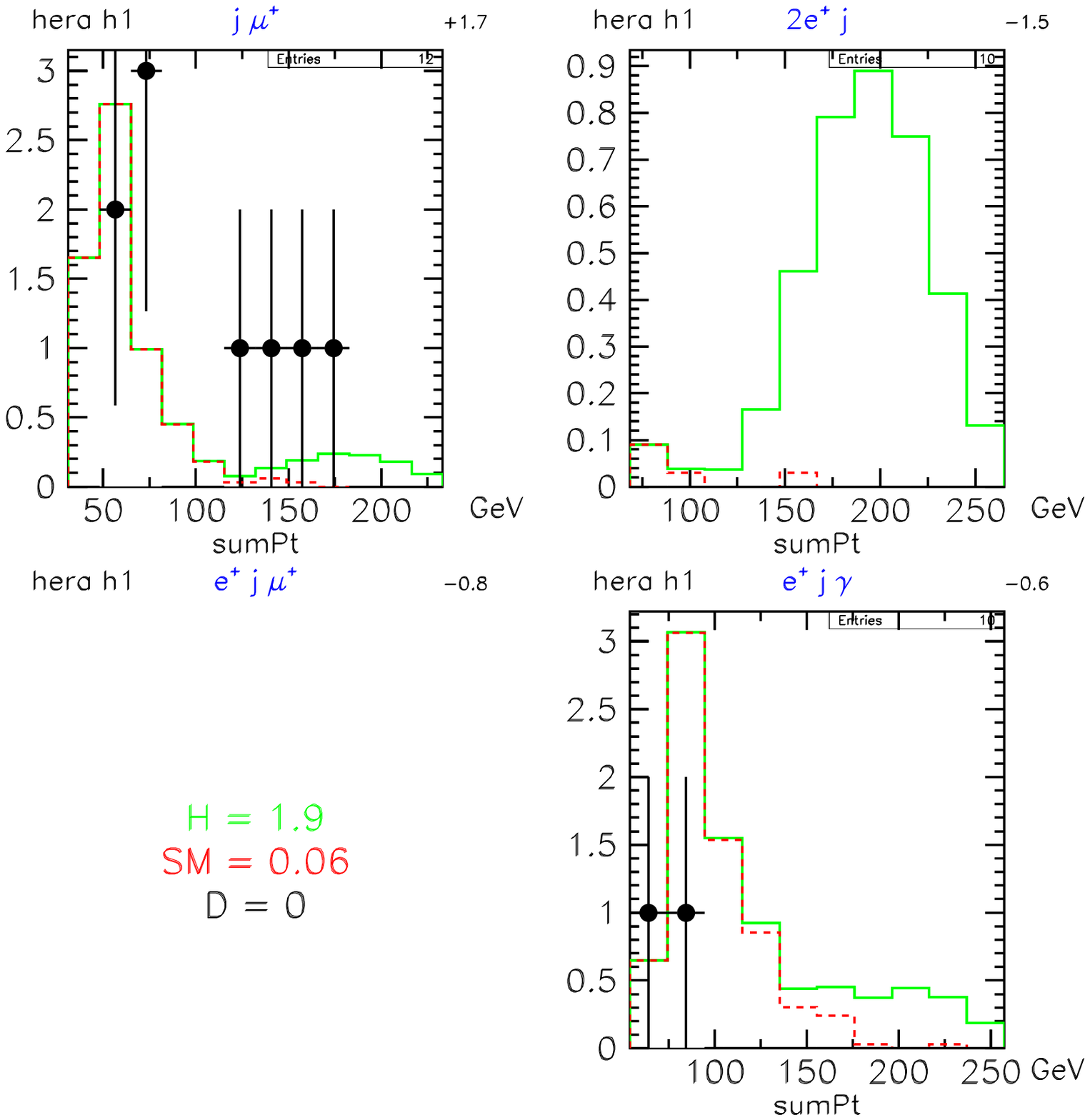}
\caption{Plots illustrating \Quaero's testing of the excited quark scenario described in the text, with parameters $f/\Lambda=0.01$ and $m_{q^*}=200$~GeV.  The contribution of the excited quark signal is seen by comparing the distributions predicted by the hypothesis \H\ (light, green) to the Standard Model \SM\ (dark, red).  \Experiment\ data are shown as (black) filled circles.  The most useful final state for distinguishing between the excited quark hypothesis \H\ and the Standard Model \SM\ is found to be $j\mu^+$, followed in importance by $2e^+ j$, $e^+j\mu^+$, and $e^+j\gamma$.  The most important variable chosen by \Quaero\ in each final state is 
shown; in the final state $e^+j\mu^+$, \Quaero\ decides it has insufficient Monte Carlo events to adequately populate a variable space, and simply considers the total number of events.
The number (black) at upper right is the decimal log likelihood ratio $\log_{10}{\L}$ that would be found by \Quaero\ if that final state were analyzed alone, without integration over systematic errors.  Systematic errors are incorporated into the final result returned by \Quaero.}
\label{fig:h1_excitedQuark_returnedPlots}
\end{figure}

\begin{figure}
\href{http://mit.fnal.gov/Quaero/quaero/doc/examples/clickableExclusionPlots/h1_excitedQuark_1.html}
{\includegraphics[width=\figuresize]{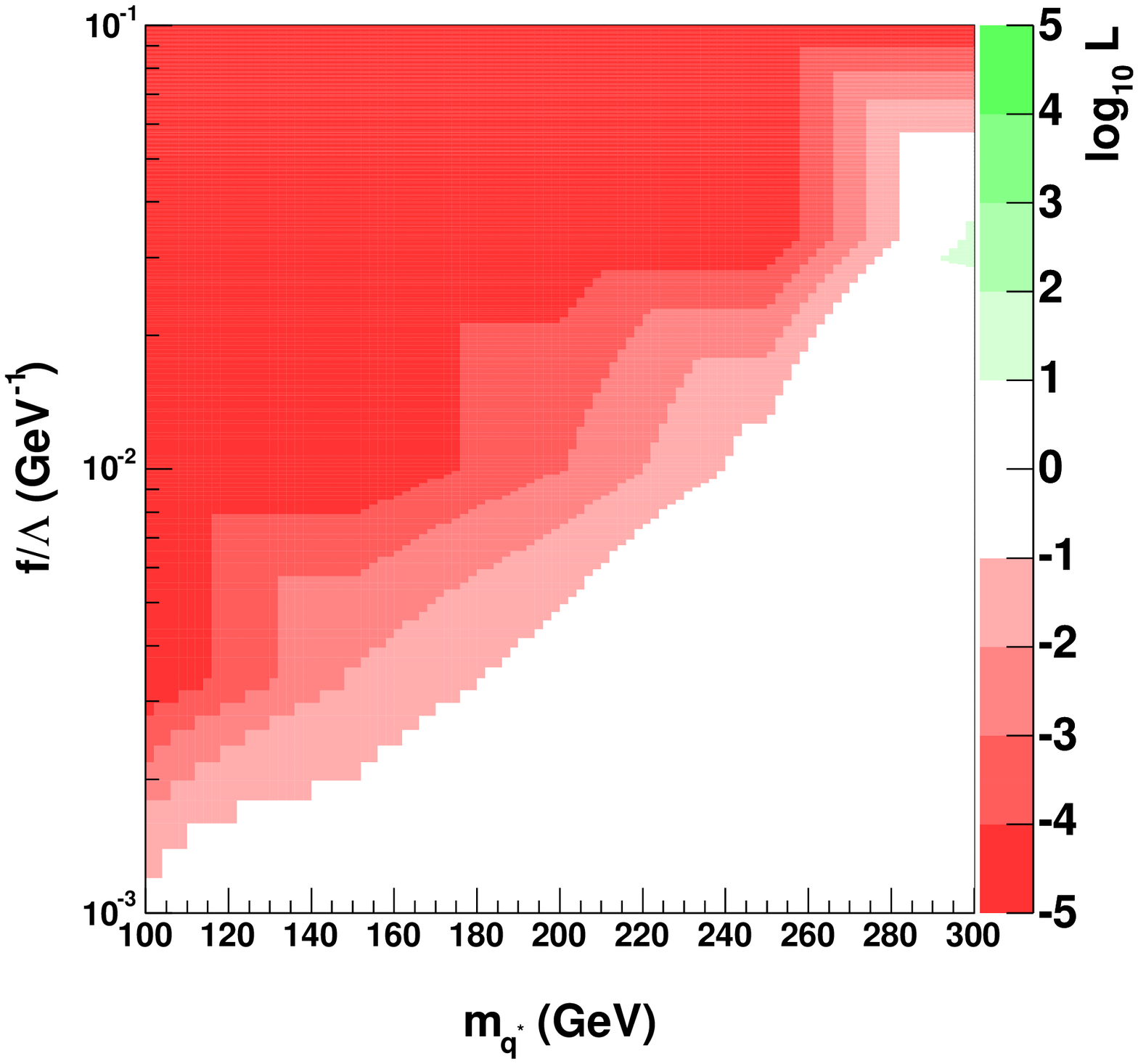}}
\caption{\Quaero's log likelihood ratio as a function of $f/\Lambda$ and $m_{q^*}$, under the assumptions $m_{u^*}=m_{d^*}$, $f=f'$, and $f_s=0$.  All shaded area corresponds to $\log_{10}{L}<0$.  
\isClickable{This exclusion plot is clickable, providing access to \Quaero's analysis of each parameter point.}
}
\label{fig:h1_excitedQuark_1_logL_2d}
\end{figure}

\begin{figure}
{\includegraphics[width=\figuresize]{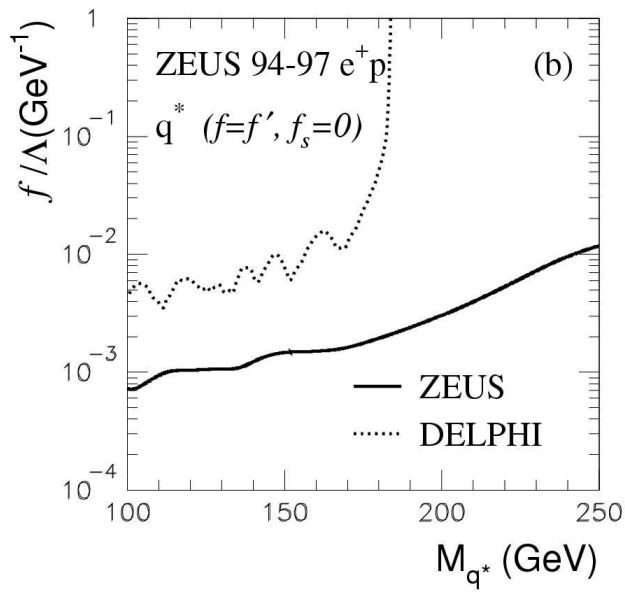}}
\caption{A previous result from {\sc{Zeus}} in the parameter plane of $f/\Lambda$ and $m_{q^*}$, under the same simplifying assumptions $f=f'$, $f_s=0$, and $m_{u^*}=m_{d^*}$, from Ref.~\cite{previousZeusExcitedQuarkAnalysis}.}
\label{fig:zeus_excitedQuark_previousResult}
\end{figure}

\begin{figure}
\begin{tabular}{cc}
\href{http://mit.fnal.gov/Quaero/quaero/doc/examples/clickableExclusionPlots/h1_excitedQuark_2.html}
{\includegraphics[width=1.5in]{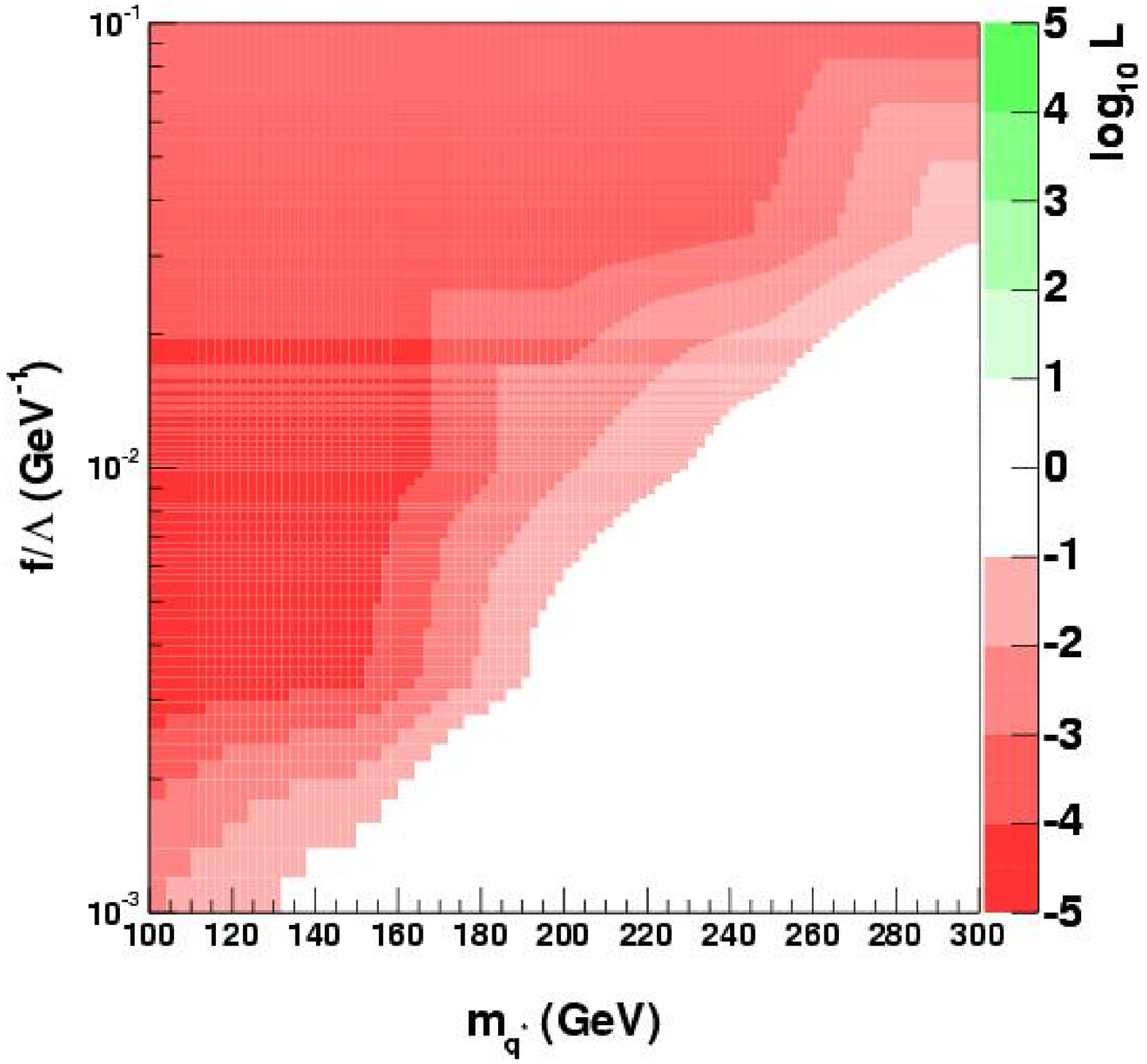}}
&
\href{http://mit.fnal.gov/Quaero/quaero/doc/examples/clickableExclusionPlots/h1_excitedQuark_3.html}
{\includegraphics[width=1.5in]{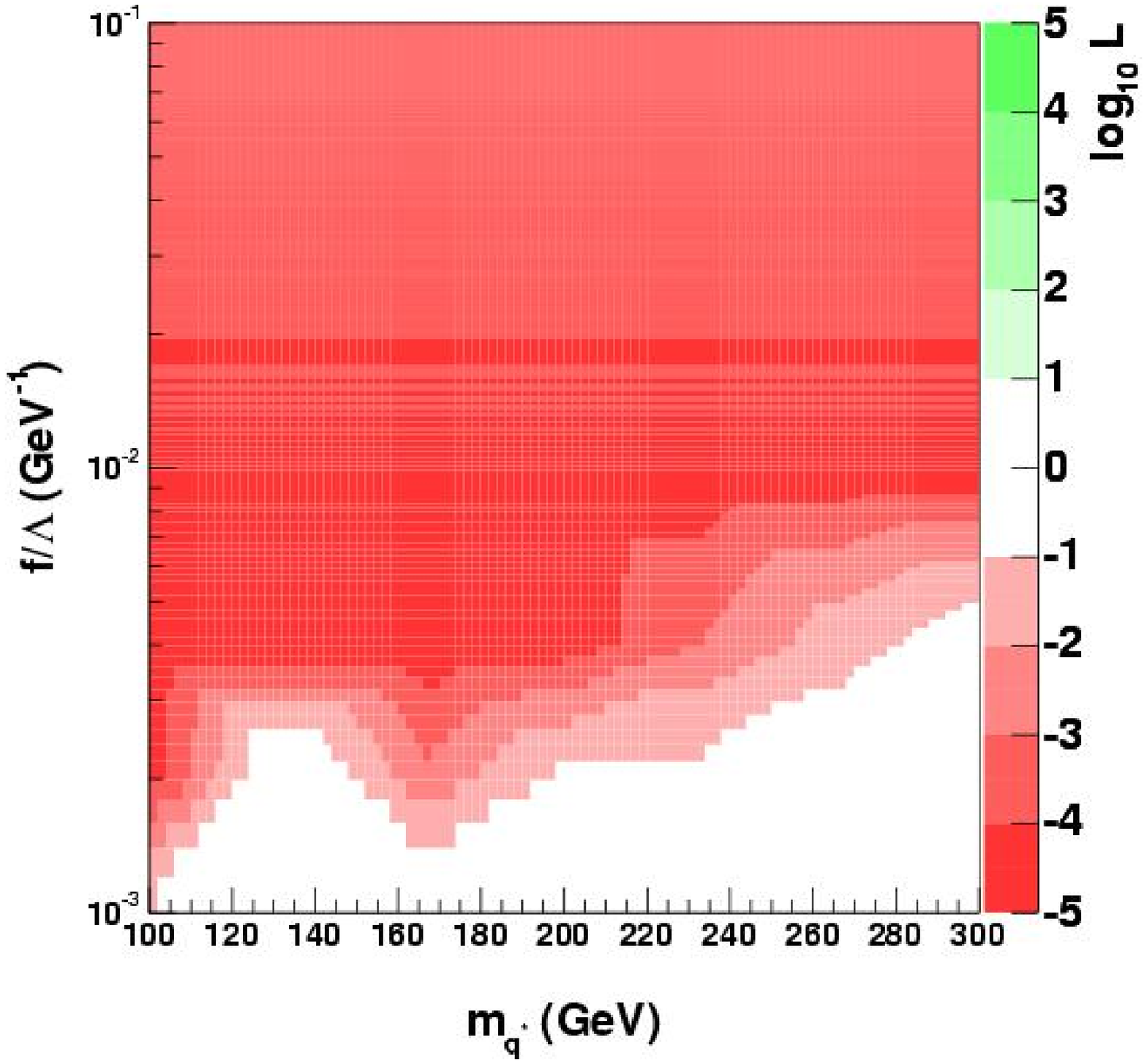}}
\\
\end{tabular}
\caption{\Quaero's log likelihood ratio as a function of $f/\Lambda$ and $m_{q^*}$, under the assumptions $m_{u^*}=m_{d^*}$, $f=-f'$, and $f_s=0$ (left), and under the assumptions $m_{q^*} \equiv m_{u^*}=2 m_{d^*}$, $f=f'$, and $f_s=0$ (right).  All shaded area corresponds to $\log_{10}{L}<0$.  
\isClickable{These exclusion plots are clickable, providing access to \Quaero's analysis of each parameter point.}
}
\label{fig:h1_excitedQuark_23_logL_2d}
\end{figure}

\subsection{Excited quark}
\label{sec:excitedQuark}

If the known fermions are composite, a clear signal would be the detection of their excited states.  At HERA, the hypothesis of an excited quark can be taken to correspond to new terms in an interaction Lagrangian of the form~\cite{ExcitedQuark:Baur:1989kv}
\begin{eqnarray}
\label{eqn:excitedQuarkLagrangian}
{\cal{L}} = & \frac{1}{2\Lambda} \bar{q}^*_R \sigma^{\mu\nu} \left( g_s f_s \frac{\lambda^a}{2} G^a_{\mu\nu} + g f \frac{\vec{\tau}}{2} \cdot \overrightarrow{W}_{\mu\nu} + g' f' \frac{Y}{2} B_{\mu\nu} \right) q_L \nonumber \\
            & + {\mbox {h.c.}},
\end{eqnarray}
where $\sigma^{\mu \nu} = (i/2) \left[ \gamma^{\mu}, \gamma^{\nu} \right]$; $g_s$, $g$, and $g'$ are the $SU(3)$, $SU(2)$, and $U(1)$ gauge couplings; $G^a_{\mu\nu}$, $\overrightarrow{W}_{\mu\nu}$, and $B_{\mu\nu}$ are the $SU(3)$, $SU(2)$, and $U(1)$ field strength tensors; $\Lambda$ is the compositeness scale; $q$ represents the first generation quark doublet; and $q^*$ is the first generation excited quark doublet.  These terms, with a specific choice of parameter values, can be specified with \MadEvent\ input shown in Fig.~\ref{fig:MadEventInput_excitedQuark}.  The choice $f=f'$, $f_s=0$, and $m_{u^*}=m_{d^*}$ leaves just two parameters:  $f/\Lambda$ and the excited quark mass $m_{q^*}$.

Parameters specified in Fig.~\ref{fig:MadEventInput_excitedQuark} include the mass of the $u^*$, its coupling to its Standard Model counterpart and photon in the first two {\tt INTERACTION} lines in Fig.~\ref{fig:MadEventInput_excitedQuark}, its coupling to its unexcited weak isospin partner and $W$ in the third and fourth {\tt INTERACTION} lines, its coupling to its Standard Model counterpart and $Z$ in the fifth and sixth {\tt INTERACTION} lines, its coupling to the photon in the seventh {\tt INTERACTION} line, and its coupling to the $Z$ boson in the eighth and final {\tt INTERACTION} line.  An excited down quark $d^*$ is introduced similarly.

Figure~\ref{fig:h1_excitedQuark_returnedPlots} summarizes \Quaero's analysis of this hypothesis with the choice $f/\Lambda=0.01$ and $m_{q^*}=200$~GeV.  Shown are the four final states contributing most to the final result.  

\Quaero's result $\log_{10}{\L}$ as a function of assumed coupling $f/\Lambda$ and excited quark mass $m_{q^*}$ is shown in Fig.~\ref{fig:h1_excitedQuark_1_logL_2d}.  The current result from {\sc{Zeus}}~\cite{previousZeusExcitedQuarkAnalysis}, which makes use of 47.7~pb$^{-1}$ of $e^+p$ data at 300~GeV, is shown in Fig.~\ref{fig:zeus_excitedQuark_previousResult}.

\Quaero\ can also be used to test other parameter points within the parameter space defined by $f$, $f'$, $f_s$, $m_{u^*}$, $m_{d^*}$, and $\Lambda$.  Fig.~\ref{fig:h1_excitedQuark_23_logL_2d}(a) shows the parameter plane of $f/\Lambda$ and $m_{q^*}$, under the assumptions $m_{u^*}=m_{d^*}$, $f=-f'$, and $f_s=0$.  Fig.~\ref{fig:h1_excitedQuark_23_logL_2d}(b) shows the parameter plane of $f/\Lambda$ and $m_{u^*}$, under the assumptions $m_{q^*} \equiv m_{u^*}=2 m_{d^*}$, $f=f'$, and $f_s=0$.  All parameter points considered are found to have $\log_{10}{L} \leq 0$.

\section{Summary}
\label{sec:Summary}

The histograms of the invariant masses and the sum of transverse momenta from the high-$p_T$ events selected in a general search for new physics at H1 have been incorporated into \Quaero, a framework for automating tests of hypotheses against data. The resulting interface is called \Quaero@\Experiment.
New physics scenarios can be provided to this interface in the form of commands to one of several commonly used event generators, which evaluate the short-distance consequences of each scenario.  
An automated analysis algorithm within \Quaero\ optimizes selection to distinguish between the new scenario and the Standard Model, returning a single number quantifying the extent to which the data (dis)favor the new hypothesis relative to the Standard Model alone, together with plots allowing the user to understand how the analysis has been performed.  
The use of \Quaero@\Experiment\ has been illustrated with searches for leptoquarks, R-parity violating supersymmetry, and excited quarks.


\acknowledgments

\Quaero\ is a simple framework for turning an understanding of collider data into statements about the underlying physics.  Many \Experiment\ physicists are responsible for the achievement of this understanding.  The authors would like to thank the Aachen and Marseille groups in particular for their roles in the \Experiment\ general search.  Emmanuelle Perez, Cristi Diaconu, and Andr\'e Sch\"oning provided guidance and review within H1.  Eckhard Elsen, Max Klein, Albrecht Wagner, and Rolf-Dieter Heuer have been sources of helpful guidance and comments within the wider DESY community.  Stephen Mrenna and Tim Stelzer provided valuable assistance with the event generators that have been incorporated within \Quaero.

Sascha Caron acknowledges the support of a Marie Curie Intra-European Fellowship within the 6th European Community Framework Programme.  Bruce Knuteson acknowledges support from a Department of Defense National Defense Science and Engineering Graduate Fellowship at the University of California, Berkeley; an International Research Fellowship from the National Science Foundation (INT-0107322); a Fermi/McCormick Fellowship at the University of Chicago; and Department of Energy grant DE-FC02-94ER40818.

\bibliography{quaero_h1}

\end{document}